\documentclass[showpacs, aps, prb, unsortedaddress,showkeys,longbibliography]{revtex4-1}

\usepackage{amssymb}
\usepackage{latexsym}
\usepackage{amsmath}
\usepackage{graphicx}
\usepackage{float}
\usepackage{xcolor}

\begin{document}

\title{On the Properties of Phononic Eigenvalue Problems}

\date{\today}
\author{Amir Ashkan Mokhtari}
\author{Yan Lu}
\author{Ankit Srivastava}
\thanks{Corresponding Author}
\affiliation{Department of Mechanical, Materials, and Aerospace Engineering,
Illinois Institute of Technology, Chicago, IL, 60616
USA}
\email{asriva13@iit.edu}

\begin{abstract}
In this paper, we consider the operator properties of various phononic eigenvalue problems. We aim to answer some fundamental questions about the eigenvalues and eigenvectors of phononic operators. These include questions about the potential real and complex nature of the eigenvalues, whether the eigenvectors form a complete basis, what are the right orthogonality relationships, and how to create a complete basis when none may exist at the outset. In doing so we present a unified understanding of the properties of the phononic eigenvalues and eigenvectors which would emerge from any numerical method employed to compute such quantities. We show that the phononic problem can be cast into linear eigenvalue forms from which such quantities as frequencies, wavenumbers, and desired components of wavevectors can be directly ascertained without resorting to searches or quadratic eigenvalue problems and that the relevant properties of such quantities can be determined apriori through the analysis of the associated operators. We further show how the Plane Wave Expansion (PWE) method may be extended to solve each of these eigenvalue forms, thus extending the applicability of the PWE method to cases beyond those which have been considered till now. The theoretical discussions are supplemented with supporting numerical calculations. The techniques and results presented here directly apply to wave propagation in other periodic systems such as photonics.

\end{abstract}

\maketitle

\section{Introduction}
There has been considerable recent research interest in the field of wave propagation in periodic structures under the fields of photonics, phononics, and even metamaterials. Much of the progress in these fields depends upon the determination of wave propagation characteristics in such periodic systems\cite{nemat2015anti,shmuel2016universality,chen2016modulating,MOKHTARI2019256}. Historically, numerical efforts in this direction have been driven towards the calculation of the so-called bandstructure\cite{2013metamaterial} of the periodic system which is a graphical representation of the frequency-wavevector pairs which satisfy a certain kind of dispersion relationship for the system. Traditionally, such dispersion relations have been calculated through what we would call the conventional form of the eigenvalue problem -- determining acceptable frequencies given a wavevector -- termed $\omega(\boldsymbol{\beta})$ systems. A host of numerical techniques have been devised to solve this particular form of the eigenvalue problem. This includes the Plane Wave Expansion (PWE) method \cite{Ho1990,2013metamaterial,kushwaha1994theory}, the multiple
scattering method\cite{Mei2005, Kafesaki1999}, variational techniques\cite{srivastava2014mixed,lu2016variational,Lu2017}, FEM \cite{white1989finite, veres2012complexity, hladky1991analysis,hussein2009reduced}, and Finite Difference \cite{Tanaka2000, chan1995order} techniques among others.

Of late, there has been growing interest in the solution of the eigenvalue problem of periodic systems when the problem is not in a traditional form, often involving complex frequencies and wavenumbers. Most commonly, complex wavenumbers and frequencies are a direct outcome of including dissipation in the system, with a complex frequency representing a temporally dissipating wave whereas a complex wavenumber representing a spatially dissipating wave. However, imaginary and/or complex wavenumbers appear even in the absence of dissipation and serve the very important purpose of satisfying interface continuity conditions in scattering problems. The simplest of the non-standard eigenvalue cases is the solution of the $\omega(\boldsymbol{\beta})$ eigenvalue form in the presence of dissipation\cite{Hussein2009,Hussein2010,Frazier2015}. In this case, it turns out that the resulting frequencies for real assumed wavevectors are complex. A further complication which has been considered in literature is the determination of the wavenumber when frequency is given (termed $\beta(\omega,\mathbf{n})$ form where $\mathbf{n}$ is the direction of wave propagation). The corresponding eigenvalue problem is most naturally quadratic and, therefore, more difficult to solve than linear eigenvalue problems. In Ref. \cite{Frazier2016}, an algorithm for 1D systems was developed that provided dispersion curves for damped free wave motion based on frequencies and wavenumbers that are permitted to be simultaneously complex. The algorithm was applied to a viscously damped mass-in-mass metamaterial exhibiting local resonance. In their study, two eigenvalue problems were solved: Frequency solutions from linear eigenvalue problem, and wavenumber solutions from quadratic eigenvalue problem. As the latter problem is quadratic, a search algorithm was presented to find the wavenumber solutions for a given frequency. The problem can alternatively be converted into a linear eigenvalue form by using a state space representation\cite{Laude2009,Moiseyenko2011,Andreassen2013,Krushynska2016}. The resulting mixed-form of the elastodynamics problem has been considered in detail in the Finite Element literature (Least-Squares FEM\cite{Gunzburger2009}) but its appearance in the phononics/photonics area is rare. Computational techniques used to solve such a problem in the area of phononics/photonics are, therefore, limited as well to what is called the Extended PWE method\cite{Hsue2005,Laude2009}. 

A further complication, rarely studied till now, which could be considered is the determination of one component of the wavevector when the other components and the frequency are given - termed $\beta_3(\omega,\beta_\alpha)$ problems. Such problems naturally emerge in cases where scattering at an interface are being studied \cite{srivastava2017evanescent,Hsue2005} This is due to the fact that Snell's law ensures that the components of the wavevector tangential to the interface are preserved and, therefore, there is a natural requirement to determine the remaining component when the preserved component is specified. Currently, there appears to exist no study in phononics which could directly solve this problem.

In this paper, we ask some basic questions pertaining to the phononic eigenvalue problems and propose some solutions. Foremost, is our attempt to analyze the phononic eigenvalue problems in the three different forms mentioned above through the lens of linear operators. This exercise reveals to us the basic properties of the eigenvalues and eigenvectors which can be expected from numerical calculations without actually doing any calculations. In all the eigenvalue forms, we are interested in determining the appropriate orthogonality conditions and whether the eigenvector basis is complete. In addition to the theoretical considerations, we also present extensions to the PWE method and give representative solutions for the eigenvalue forms considered. This paper is organized as follows: in section II a brief introduction of the spectral theorem and properties of self-adjoint and non-self-adjoint operators are presented. We then investigate the properties of $\omega(\beta)$ problem for different cases of material properties and also different types of wave number (real and complex) in section III. In section IV, the general problem of $\beta(\omega)$ and the corresponding properties of its eigenvalues and eigenvectors are studied. One special case of this problem is the $\beta_3(\omega,\beta_{\alpha})$ problem which is formulated in section IV.

\section{Some results on the eigenvalue properties of linear operators}
\noindent Here we are concerned with eigenvalue  problems of the following form:
\begin{eqnarray}\label{eq:GeneralizedEigProb}
\displaystyle Av=\lambda Bv
\end{eqnarray}
where $A,B$ are linear operators which act on a dense domain in the Hilbert space $H$ of complex functions over which an inner product has been defined \cite{Schmdgen2012,Davies1995}:
\begin{eqnarray}
\displaystyle \langle u,v\rangle=\int u v^* dx
\end{eqnarray}
where $u,v\in H$. The adjoint operator to $A$, denoted by $A^*$, is defined by:
\begin{eqnarray}
\displaystyle \langle Au,v\rangle=\langle u,A^*v\rangle
\end{eqnarray}
Consider the case where $A,B$ are normal operators:
\begin{eqnarray}
AA^*=A^*A;\quad \vert Av\vert=\vert A^*v\vert
\end{eqnarray}
with similar relations for $B$. If $\lambda$ is an eigenvalue of $A$ with associated eigenvector $v$ ($ Av=\lambda Bv$) then
\begin{eqnarray}
\vert(\displaystyle A-\lambda B)v\vert=\vert(\displaystyle A-\lambda B)^*v\vert=0,
\end{eqnarray}
showing that $\lambda^*$ is an eigenvalue of $A^*$ with the same eigenvector ($A^*v=\lambda^* B^*v$). If $\lambda_1,\lambda_2$ are two distinct eigenvalues with associated eigenvectors $v,w$ then
\begin{eqnarray}\label{eq:orthogonalNormaloperator}
\lambda_1\langle Bv,w\rangle=\langle Av,w\rangle=\langle v,A^*w\rangle=\langle v,\lambda_2^*B^*w\rangle=\lambda_2\langle Bv,w\rangle
\end{eqnarray}
showing that the eigenvectors are orthogonal ($\langle Bv,w\rangle=0$). 

Most of the standard eigenvalue problems in phononics and photonics and other areas of physics are governed by linear operators that are self-adjoint or Hermitian in character\cite{Kostenbauder1997}. The normal modes of such a system, referring to the eigenfunctions of the Hermitian operator, then form a complete and orthonormal set for expanding the field variables in the system like displacement, stress, etc. \cite{Morse1953}. However, there are several important physical problems in phononics (like scattering of the wave at the metamaterial interfaces) and other ﬁelds, for which the governing operator is not Hermitian or self-adjoint and cannot easily be made so. 
There is no universal way for unbounded differential operators to figure out if their eigenfunctions form a complete basis or not. But at a very general level, if a differential operator is acting on a finite domain with periodic boundary condition, then it has a point spectrum and it is possible to expand the state of the system using the normal eigenmodes \cite{Davies1995}. If the operator is not normal, it still has propagation eigenmode. However, these modes are no longer orthogonal. In fact, they are orthogonal to the eigenmodes of the adjoint operator, a case which is called bi-orthogonality\cite{Kostenbauder1997}.  Below we present a brief explanation on the spectral theory for Hermitian operators, non-Hermitian operators, the problem of bi-orthogonality and also the generalized eigenvectors and formation of a complete basis.

\smallskip
\noindent \paragraph{Hermitian operators:} Self-adjoint operators ($A=A^*$) form a subset of normal operators and if $A,B$ are self-adjoint and $B$ is positive definite then the associated eigenvalues are real. In both cases the eigenvectors form a complete basis. The matrix representations of $A,B$ are $[A],[B]$ respectively. If $A$ is positive definite ($\langle Av,v\rangle>0$) then $[A]$ is also positive definite ($\{x\}^\dagger [A] \{x\}>0$) where $\dagger$ represents conjugate transpose. Similar consideration holds for negative-definiteness as well. Using a finite set of basis functions, the generalized eigenvalue problem can be represented in its finite dimensional form:
\begin{eqnarray}
\displaystyle [A]\{v\}=\lambda [B]\{v\}
\end{eqnarray}
Now if $[B]$ is positive definite then all its eigenvalues are positive. Since the eigenvalues of $[B]^{-1}$ are the inverse of the eigenvalues of $[B]$, its eigenvalues are also positive. Therefore, positive-definiteness of $[B]$ implies the positive-definiteness of $[B]^{-1}$. Now converting the generalized eigenvalue problem above into a standard eigenvalue problem:
\begin{eqnarray}
\displaystyle [C]\{v\}=\lambda \{v\};\quad [C]=[B]^{-1}[A]
\end{eqnarray}
Denoting $[B]^{-1}=[H]$ we have $(\{v\}[A])^\dagger [H][A]\{v\}=\{v\}^\dagger[A] [H][A]\{v\}>0$ since $[H]$ is positive definite and $[A]$ is Hermitian. Now we have the generalized eigenvalue problem
\begin{eqnarray}
\nonumber \displaystyle [H][A]\{v\}=\lambda \{v\}\\
\nonumber \displaystyle \{v\}^\dagger[A][H][A]\{v\}=\lambda \{v\}^\dagger[A]\{v\}\\
\displaystyle \lambda=\frac{\{v\}^\dagger[A][H][A]\{v\}}{\{v\}^\dagger[A]\{v\}}
\end{eqnarray}
showing that all eigenvalues $\lambda$ of the generalized eigenvalue problem are negative since the numerator is positive and the denominator is negative. Therefore, in a eigenvalue problem $Av=\lambda Bv$ where $A$ is negative definite and self-adjoint and $B$ is positive definite, all $\lambda$ are real and negative.

\smallskip
\noindent \paragraph{Non-Hermitian Operators:}
Non-Hermitian systems have drawn significant recent attention due to their intriguing physics revolving $\mathcal{PT}-$symmetry \cite{bender1998real} and exceptional points \cite{heiss2000repulsion,heiss1990avoided,lu2018level}. A recent review on non-Hermitian physics is given by Ref \cite{el2018non}. Questions about orthogonality condition and completeness of basis of non-Hermitian operators are of particular interest in wave scattering settings because they require an appropriate set of wave functions to expand wave fields in \cite{srivastava2016metamaterial}. In connection to the earlier results, any self-adjoint (Hermitian) operator can be shown to be normal. Vice-versa, if an operator is not normal then it can be shown to be non-self-adjoint (non-Hermitian). It is well known that the right eigenvectors of a non-Hermitian operator are not orthogonal and an orthogonalization process such as the Gram-Schmidt method does not guarantee to work unless the eigenvectors are at least linearly independent \cite{arfken1985mathematical}. 
To discuss this further, we need to consider two cases, diagonalizable and non-diagonalizable operators. In the discussions below, the square brackets denoting matrices are dropped and the matrix nature of the equations is assumed implicit.

\smallskip
\noindent First, for a generalized eigenvalue problem such as Eq. (\ref{eq:GeneralizedEigProb}) we consider its left eigenvector $u_l$ which satisfies
\begin{equation}
u_l^\dagger A=\lambda u_l^\dagger B
\end{equation}
Note that $u_l$ can be solved as the right eigenvector of $A^*$ and $B^*$ with associated eigenvalue $\lambda^*$. If $\lambda_1$ is an eigenvalue with left eigenvector $u_l^\dagger$ and $\lambda_2\neq\lambda_1$ is another eigenvalue with right eigenvector $u_r$ then we have:
\begin{equation}
\lambda_1u_l^\dagger Bv_r=u_l^\dagger Av_r=\lambda_2 u_l^\dagger Bv_r,
\end{equation}
The above shows that for two distinct eigenvalues $\lambda_1,\lambda_2$ of a non-Hermitian eigenvalue problem, the bi-orthogonality relation $u_l^\dagger Bv_r=0$ is satisfied. In operator notation, this bi-orthogonality condition can be written as $\langle B(\mathbf{v_r}),\mathbf{u_l}\rangle=0$. A numerical example which verifies this bi-orthogonality for a relevant phononic case will be given later in Section \ref{subsection:beta3PWE}.

This bi-orthogonality condition for non-Hermitian operators has been widely accepted in the physics community \cite{el2018non}, however, it should be noted that it only holds under the assumption that the eigenvalues are distinct (operator is diagonalizable). If the non-Hermitian operator has repeated eigenvalues (such as in the case of exceptional points), then it can be shown to be non-diagonalizable\cite{riechers2018beyond} in which case, the bi-orthogonality condition will not apply and enough linearly independent eigenvectors cannot be found through conventional eigenvalue computation. However, in such a case, the operator can be transformed into its Jordan canonical form, and generalized left and right eigenvectors corresponding to each Jordan block can be defined which then satisfy a bi-orthogonality condition. The generalized eigenvectors thus calculated will form a linearly independent set.

For example, if a repeated eigenvalue $\lambda$ corresponds to a Jordan block, $\mathbf{J}_i$, of size $m_i$, then it can be associated with one pair of conventional left and right eigenvectors $u_l^{(1)}$ and $u_r^{(1)}$, and $m_i-1$ pairs of generalized left and right eigenvectors defined through the modified eigenvalue equation
\begin{eqnarray}
    (A-\lambda B)^mu_r^{(m)}=0;\quad u_l^{(m)\dagger}(A-\lambda B)^m=0,
\end{eqnarray}
where $0\leq m\leq m_i-1$ with $u_r^{(0)}=0$ and $u_l^{(0)}=0$. The generalized eigenvectors are obtained by first solving the conventional eigenvectors and then following the process shown below
\begin{eqnarray}
    \label{eq:rightGenEigVec}
    (A-\lambda B)  u_r^{(m+1)}=Bu_r^{(m)}\\
    \label{eq:leftGenEigVec}
    u_l^{(m+1)\dagger}  (A-\lambda B)=u_l^{(m)\dagger}B.
\end{eqnarray}
Ref. \cite{riechers2018beyond} has presented an excellent discussion on the bi-orthogonality properties of generalized eigenvectors. Here, we only note the primary result of immediate relevance to us. If $u_{lj}^{(m)},u_{rj}^{(n)}$ ($u_{lk}^{(m)},u_{rk}^{(n)}$) are the left and right generalized eigenvectors corresponding to the repeated eigenvalue $\lambda_j$ ($\lambda_k$) for which there exists the Jordan block $\mathbf{J}_j$ ($\mathbf{J}_k$) of size $m_j$ ($m_k$) then the following general bi-orthogonality relationship holds:
\begin{equation}
    u_{lj}^{(n)\dagger} B u_{rk}^{(m)}=\delta_{jk}\delta_{m+n,m_i+1}.
\end{equation}
This completes the discussion of the orthogonality conditions for non-Hermitian operators. It must be noted that for phononic eigenvalue problems, the need to create generalized left and right eigenvectors will only be in the case where an exceptional points is involved in the eigen-spectrum. In the sections below, we consider the phononic eigenvalue problem under various operator forms, elucidating their relevant properties. We also present simultaneous numerical examples which explicitly show how the operator equations may be solved. These numerical solutions are based upon modifications of the well studied Plane Wave Expansion method, however, other numerical methods are also possible. The theoretical discussions are presented in the form of stiffness tensors at some places and compliance tensors at other places. The choice is mainly directed by the simplicity of resulting expressions and has no bearing on the conclusions reached. The PWE scheme is always implemented based upon the stiffness tensor in this paper.

\section{$\omega(\beta)$ solutions for a wave with a given wavevector}
Consider Bloch waves propagating in a phononic crystal in direction $\mathbf{n}$. The displacement and stress fields due to the wave will have the general form:
\begin{eqnarray}
\displaystyle \nonumber \mathbf{u}=\bar{\mathbf{u}}\exp\left[i(\omega t-\beta\mathbf{n}\cdot\mathbf{x})\right]\\
\displaystyle \boldsymbol{\sigma}=\bar{\boldsymbol{\sigma}}\exp\left[i(\omega t-\beta\mathbf{n}\cdot\mathbf{x})\right]
\end{eqnarray}
where $\bar{\mathbf{u}},\bar{\boldsymbol{\sigma}}$ are $\Omega-$periodic. $\omega$ and $\beta$ can both potentially assume real, imaginary, or complex values, however, only certain combinations are physically meaningful. First consider the usual form of the eigenvalue problem:
\begin{eqnarray}
\displaystyle (C_{ijkl}\bar{u}_{k,l}-i\beta n_lC_{ijkl}\bar{u}_k)_{,j}-i\beta n_jC_{ijkl}\bar{u}_{k,l}-\beta^2n_jC_{ijkl}n_l\bar{u}_k=\lambda\rho \bar{u}_i
\end{eqnarray}
where $\lambda=-\omega^2$. which is in the form
\begin{eqnarray}
\displaystyle A\bar{v}=\lambda B\bar{v}
\end{eqnarray}
when one identifies $\bar{v}\equiv \{\bar{\mathbf{u}}\}$. Now consider two $\Omega$ periodic fields $\bar{v},\bar{w}$. The relevant inner product for the operator $A$ is:
\begin{eqnarray}
\displaystyle \nonumber \langle A\bar{v},\bar{w}\rangle=\int\left[\displaystyle (C_{ijkl}\bar{v}_{k,l}-i\beta n_lC_{ijkl}\bar{v}_k)_{,j}-i\beta n_jC_{ijkl}\bar{v}_{k,l}-\beta^2n_jC_{ijkl}n_l\bar{v}_k\right]\bar{w}_i^*\mathrm{d}\Omega
\end{eqnarray}
Since $C_{ijkl},\bar{v},\bar{w}$ are all $\Omega$ periodic terms, the above can be transformed using Gauss theorem into:
\begin{eqnarray}
\displaystyle \nonumber \langle A\bar{v},\bar{w}\rangle=\int\displaystyle \bar{v}_k\left[(C_{ijkl}\bar{w}_{i,j}^*+i\beta n_jC_{ijkl}\bar{w}_i^*)_{,l}+i\beta n_lC_{ijkl}\bar{w}_{i,j}^*-\beta^2n_jC_{ijkl}n_l\bar{w}_i^*\right]\mathrm{d}\Omega\equiv \langle \bar{v},A^*\bar{w}\rangle
\end{eqnarray}
showing that the adjoint operator $A^*$ is:
\begin{eqnarray}
\displaystyle A^*\bar{w}=(C_{ijkl}^*\bar{w}_{i,j}-i\beta^* n_jC_{ijkl}^*\bar{w}_i)_{,l}-i\beta^* n_lC_{ijkl}^*\bar{w}_{i,j}-\beta^{*2}n_jC_{ijkl}^*n_l\bar{w}_i
\end{eqnarray}
In what follows we will use $\mathbf{C}$ for the tensor $C_{ijkl}$, $\mathbf{n}$ for the vector $n_k$, and $\bar{\mathbf{v}}$ for the vector $\bar{v}_k$. Tensor contraction to the right of $\mathbf{C}$ will represent contraction with respect to the last two indices and to the left will represent contraction with the the first two indices. Appropriate contractions are assumed without making them explicit. With this we have:
\begin{gather}
\label{norm_cond1}
\displaystyle \vert A\bar{v}\vert^2=\langle A\bar{v},A\bar{v}\rangle\\
=\displaystyle \nonumber \int\left[ \nabla\cdot(\mathbf{C}\nabla\bar{\mathbf{v}}-i\beta \mathbf{C}\mathbf{n}\bar{\mathbf{v}})-i\beta \mathbf{n}\mathbf{C}\nabla\bar{\mathbf{v}}-\beta^2\mathbf{n}\mathbf{C}\mathbf{n}\bar{\mathbf{v}}\right]\cdot\left[ \nabla\cdot(\mathbf{C}^*\nabla\bar{\mathbf{v}}^*+i\beta^* \mathbf{C}^*\mathbf{n}\bar{\mathbf{v}}^*)+i\beta^* \mathbf{n}\mathbf{C}^*\nabla\bar{\mathbf{v}}^*-\beta^{*2}\mathbf{n}\mathbf{C}^*\mathbf{n}\bar{\mathbf{v}}^*\right]\mathrm{d}\Omega\\
=\displaystyle \nonumber \int\left[ \nabla\cdot\mathbf{C}\nabla\bar{\mathbf{v}}-\beta^2\mathbf{n}\mathbf{C}\mathbf{n}\bar{\mathbf{v}}-i\beta (\nabla\cdot\mathbf{C}\mathbf{n}\bar{\mathbf{v}}+ \mathbf{n}\mathbf{C}\nabla\bar{\mathbf{v}})\right]\cdot\left[ \nabla\cdot\mathbf{C}^*\nabla\bar{\mathbf{v}}^*-\beta^{*2}\mathbf{n}\mathbf{C}^*\mathbf{n}\bar{\mathbf{v}}^*+i\beta^* (\nabla\cdot\mathbf{C}^*\mathbf{n}\bar{\mathbf{v}}^*+ \mathbf{n}\mathbf{C}^*\nabla\bar{\mathbf{v}}^*)\right]\mathrm{d}\Omega\\
=\displaystyle \nonumber \int\left[ \mathbf{a}\cdot\mathbf{a}^*+\vert\beta\vert^2\mathbf{b}\cdot\mathbf{b}^*-i\beta\mathbf{a}^*\cdot\mathbf{b}+i\beta^*\mathbf{a}\cdot\mathbf{b}^*\right]\mathrm{d}\Omega
\end{gather}
where $\mathbf{a}=(\nabla\cdot\mathbf{C}\nabla\bar{\mathbf{v}}-\beta^2\mathbf{n}\mathbf{C}\mathbf{n}\bar{\mathbf{v}})$ and $\mathbf{b}=(\nabla\cdot\mathbf{C}\mathbf{n}\bar{\mathbf{v}}+ \mathbf{n}\mathbf{C}\nabla\bar{\mathbf{v}})$. On the other hand we have:
\begin{gather}
\label{norm_cond2}
\displaystyle \vert A^*\bar{v}\vert^2=\langle A^*\bar{v},A^*\bar{v}\rangle\\
=\displaystyle \nonumber \int\left[ \nabla\cdot(\nabla\bar{\mathbf{v}}\mathbf{C}^*-i\beta^* \mathbf{n}\bar{\mathbf{v}}\mathbf{C}^*)-i\beta^* \nabla\bar{\mathbf{v}}\mathbf{C}^*\mathbf{n}-\beta^{*2}\mathbf{n}\bar{\mathbf{v}}\mathbf{C}^*\mathbf{n}\right]\cdot\left[ \nabla\cdot(\nabla\bar{\mathbf{v}}^*\mathbf{C}+i\beta \mathbf{n}\bar{\mathbf{v}}^*\mathbf{C})+i\beta \nabla\bar{\mathbf{v}}^*\mathbf{C}\mathbf{n}-\beta^{2}\mathbf{n}\bar{\mathbf{v}}^*\mathbf{C}\mathbf{n}\right]\mathrm{d}\Omega\\
=\displaystyle \nonumber \int\left[ \mathbf{c}\cdot\mathbf{c}^*+\vert\beta\vert^2\mathbf{d}\cdot\mathbf{d}^*-i\beta^*\mathbf{c}^*\cdot\mathbf{d}+i\beta\mathbf{c}\cdot\mathbf{d}^*\right]\mathrm{d}\Omega
\end{gather}
where $\mathbf{c}=(\nabla\cdot\nabla\bar{\mathbf{v}}\mathbf{C}^*-\beta^{*2}\mathbf{n}\bar{\mathbf{v}}\mathbf{C}^*\mathbf{n})$ and $\mathbf{d}=(\nabla\cdot\mathbf{n}\bar{\mathbf{v}}\mathbf{C}^*+ \nabla\bar{\mathbf{v}}\mathbf{C}^*\mathbf{n})$. Therefore, in general, $\vert A\bar{v}\vert\neq\vert A^*\bar{v}\vert$, or $A$ is not a normal operator. However, for certain special cases $A$ becomes a normal operator and the spectral theorem applies to it. The stiffness tensor $\mathbf{C}$ can always be separated into its Hermitian and skew-Hermitian parts $\mathbf{C}=\mathbf{H}+\mathbf{N}$ where $H^*_{klij}=H_{ijkl}$ and $N^*_{klij}=-N_{ijkl}$.
\subsection{Real $\beta$}

\paragraph{$\mathbf{C}=\mathbf{H}$:} Consider the case when the skew-Hermitian part is zero ($\mathbf{N}=0$). In this case, $C_{klij}^*=C_{ijkl}$. and we have $\mathbf{c}=(\nabla\cdot\mathbf{C}\nabla\bar{\mathbf{v}}-\beta^{*2}\mathbf{n}\mathbf{C}\mathbf{n}\bar{\mathbf{v}})$ and $\mathbf{d}=(\nabla\cdot\mathbf{C}\mathbf{n}\bar{\mathbf{v}}+ \mathbf{n}\mathbf{C}\nabla\bar{\mathbf{v}})$. If we further insist that $\beta$ is real ($\beta^*=\beta$) then $\mathbf{c}=\mathbf{a},\mathbf{d}=\mathbf{b}$ and, in fact, $\vert A\bar{v}\vert=\vert A^*\bar{v}\vert$ and $A$ is a normal operator. Moreover, in this case the adjoint operator is
\begin{eqnarray}
\displaystyle A^*\bar{w}=(C_{klij}\bar{w}_{i,j}-i\beta n_jC_{klij}\bar{w}_i)_{,l}-i\beta n_lC_{klij}\bar{w}_{i,j}-\beta^{2}n_jC_{klij}n_l\bar{w}_i
\end{eqnarray}
showing that $A^*=A$, or that $A$ is a self-adjoint operator. Since the operator $B$ is clearly self-adjoint (and, therefore, normal), the eigenvalue problem which emerges from assuming a real wavenumber in an elastodynamic system characterized by a stiffness tensor which respects the symmetry $C_{ijkl}^*=C_{klij}$ is self adjoint. The eigenvalues, $\lambda$, are, therefore, real and the corresponding frequencies $\omega=\pm\sqrt{-\lambda}$ can be either real or imaginary (but not complex with simultaneously nonzero real and imaginary parts). We further have:
\begin{gather}
\displaystyle \langle A\bar{v},\bar{v}\rangle=\int\left[\displaystyle (C_{ijkl}\bar{v}_{k,l}-i\beta n_lC_{ijkl}\bar{v}_k)_{,j}-i\beta n_jC_{ijkl}\bar{v}_{k,l}-\beta^2n_jC_{ijkl}n_l\bar{v}_k\right]\bar{v}_i^*\mathrm{d}\Omega\\
\nonumber=\int\left[\displaystyle -\bar{v}_{i,j}^*C_{ijkl}\bar{v}_{k,l}-\beta^2n_j\bar{v}_i^*C_{ijkl}n_l\bar{v}_k+i\beta \bar{v}_{i,j}^*C_{ijkl}n_l\bar{v}_k-i\beta n_j\bar{v}_i^*C_{ijkl}\bar{v}_{k,l}\right]\mathrm{d}\Omega\\
\nonumber=-\int\left[\displaystyle \bar{v}_{i,j}^*C_{ijkl}\bar{v}_{k,l}+\beta^2n_j\bar{v}_i^*C_{ijkl}n_l\bar{v}_k-i\beta \bar{v}_{i,j}^*C_{ijkl}n_l\bar{v}_k+i\beta n_j\bar{v}_i^*C_{ijkl}\bar{v}_{k,l}\right]\mathrm{d}\Omega\\
\nonumber=-\int\left[ \bar{v}_{i,j}^*-i\beta \bar{v}_i^*n_j\right]C_{ijkl}\left[ \bar{v}_{k,l}+i\beta \bar{v}_kn_l\right]\mathrm{d}\Omega
\end{gather}
which is of the form $-\int s_{ij}C_{ijkl}s_{kl}^*\mathrm{d}\Omega$ when one identifies $s_{ij}\equiv \bar{v}_{i,j}^*-i\beta \bar{v}_i^*n_j$. If we assume that the stiffness tensor is positive-definite (as in conservative systems) then the integral is always less than zero showing that the operator $A$ is negative definite as well ($\langle A\bar{v},\bar{v}\rangle\leq 0$). Since the operator $B$ is clearly positive definite, this means that all the eigenvalues $\lambda$ of the generalized eigenproblem under real $\beta$ and a self-adjoint and positive definite $\mathbf{C}$ tensor will be negative. All the frequencies $\omega=\sqrt{-\lambda}$ will, therefore, be purely real. As an academic point which is a corollary of this analysis, if $\mathbf{C}$ is negative definite then all frequencies will be imaginary since all $\lambda$ will be positive and real. 
\smallskip

\paragraph{$\mathbf{C}=\mathbf{N}$:} If the Hermitian part of the stiffness tensor is zero ($\mathbf{H}=0$) then $C^*_{klij}=-C_{ijkl}$. In this case, $\mathbf{c}=(-\nabla\cdot\mathbf{C}\nabla\bar{\mathbf{v}}+\beta^{*2}\mathbf{n}\mathbf{C}\mathbf{n}\bar{\mathbf{v}})$ and $\mathbf{d}=(-\nabla\cdot\mathbf{C}\mathbf{n}\bar{\mathbf{v}}- \mathbf{n}\mathbf{C}\nabla\bar{\mathbf{v}})$. If $\beta$ is such that $\beta^{*2}=\beta^2$ then we will have $\mathbf{c}=-\mathbf{a}$ and $\mathbf{d}=-\mathbf{b}$. Of course, in this situation we will have $\mathbf{c}\cdot\mathbf{c}^*=\mathbf{a}\cdot\mathbf{a}^*,\mathbf{d}\cdot\mathbf{d}^*=\mathbf{b}\cdot\mathbf{b}^*$. Even in this case if $\beta$ is real then it is clear that $\vert A\bar{v}\vert=\vert A^*\bar{v}\vert$ and that $A$ is a normal operator. In this case, we have:
\begin{eqnarray}
\displaystyle A^*\bar{w}=(-C_{klij}\bar{w}_{i,j}+i\beta n_jC_{klij}\bar{w}_i)_{,l}+i\beta n_lC_{klij}^*\bar{w}_{i,j}+\beta^{2}n_jC_{klij}^*n_l\bar{w}_i
\end{eqnarray}
showing that $A^*\neq A$. Therefore, unlike in the case where $\mathbf{N}=0$, $\mathbf{H}=0$ (for real $\beta$) leads to an eigenvalue problem which is not self-adjoint. It is, however, normal which means that the eigenvectors will be orthogonal and will form a complete basis. The eigenvalues, in this case, have no requirement of being real. Instead we have $A^*=-A$ and, therefore,
\begin{eqnarray}
\lambda\langle Bv,v\rangle=\langle Av,v\rangle=\langle v,A^*v\rangle=\langle v,-Av\rangle=-\lambda^*\langle v,Bv\rangle=-\lambda^*\langle Bv,v\rangle
\end{eqnarray}
showing that $\lambda^*=-\lambda$. This, in turn, means that all eigenvalues of the problem will be strictly imaginary. Since $\omega=\sqrt{-\lambda}$, the corresponding frequencies will be complex with equal nonzero real and imaginary parts. 
\smallskip
\paragraph{$\mathbf{C}=\mathbf{N}+\mathbf{H}$:} In this case, the tensors $\mathbf{a},\mathbf{b}$ can be divided into two parts: one resulting from $\mathbf{H}$ and the other resulting from $\mathbf{N}$. We can write $\mathbf{a}=\mathbf{a}^h+\mathbf{a}^n$ and similarly for $\mathbf{b}$. Carrying out the same decomposition for $\mathbf{c},\mathbf{d}$, we can show that $\mathbf{c}=\mathbf{a}^h-\mathbf{a}^n$ and $\mathbf{d}=\mathbf{b}^h-\mathbf{b}^n$. As $\mathbf{c}\cdot\mathbf{c}^*\neq\mathbf{a}\cdot\mathbf{a}^*$ and $\mathbf{d}\cdot\mathbf{d}^*\neq\mathbf{b}\cdot\mathbf{b}^*$ then $\vert A\bar{v}\vert^2 \neq \vert A^*\bar{v}\vert^2 $ and the operator $A$ is not normal. In this case, the eigenvalues $\lambda$ are complex which result in complex frequencies.

\subsection{Imaginary and complex $\beta$}
Assuming $\beta=i\beta_{\Im}$, we have $\mathbf{c}=(\nabla\cdot\mathbf{C}\nabla\bar{\mathbf{v}}+\beta_{\Im}^{2}\mathbf{n}\mathbf{C}\mathbf{n}\bar{\mathbf{v}})$, $\mathbf{d}=(\nabla\cdot\mathbf{C}\mathbf{n}\bar{\mathbf{v}}+ \mathbf{n}\mathbf{C}\nabla\bar{\mathbf{v}})$, $\mathbf{a}=(\nabla\cdot\mathbf{C}\nabla\bar{\mathbf{v}}+\beta_{\Im}^2\mathbf{n}\mathbf{C}\mathbf{n}\bar{\mathbf{v}})$, and $\mathbf{b}=(\nabla\cdot\mathbf{C}\mathbf{n}\bar{\mathbf{v}}+ \mathbf{n}\mathbf{C}\nabla\bar{\mathbf{v}})$. then $\mathbf{c}=\mathbf{a},\mathbf{d}=\mathbf{b}$, however as $i\beta^*\neq i\beta$, then from Eqs.(\ref{norm_cond1},\ref{norm_cond2}), we have  $\vert A\bar{v}\vert^2 \neq \vert A^*\bar{v}\vert^2$. When  $\beta=\beta_{\Re}+i\beta_{\Im}$, then $\mathbf{c}\neq\mathbf{a},\mathbf{d}\neq\mathbf{b}$ and consequently the operator is not normal when $\beta$ is not real. In these cases, the eigenvalues are complex.
\subsection{$\omega(\beta)$ solutions using PWE}
\begin{figure}[htp]
\centering
\includegraphics[scale=.55]{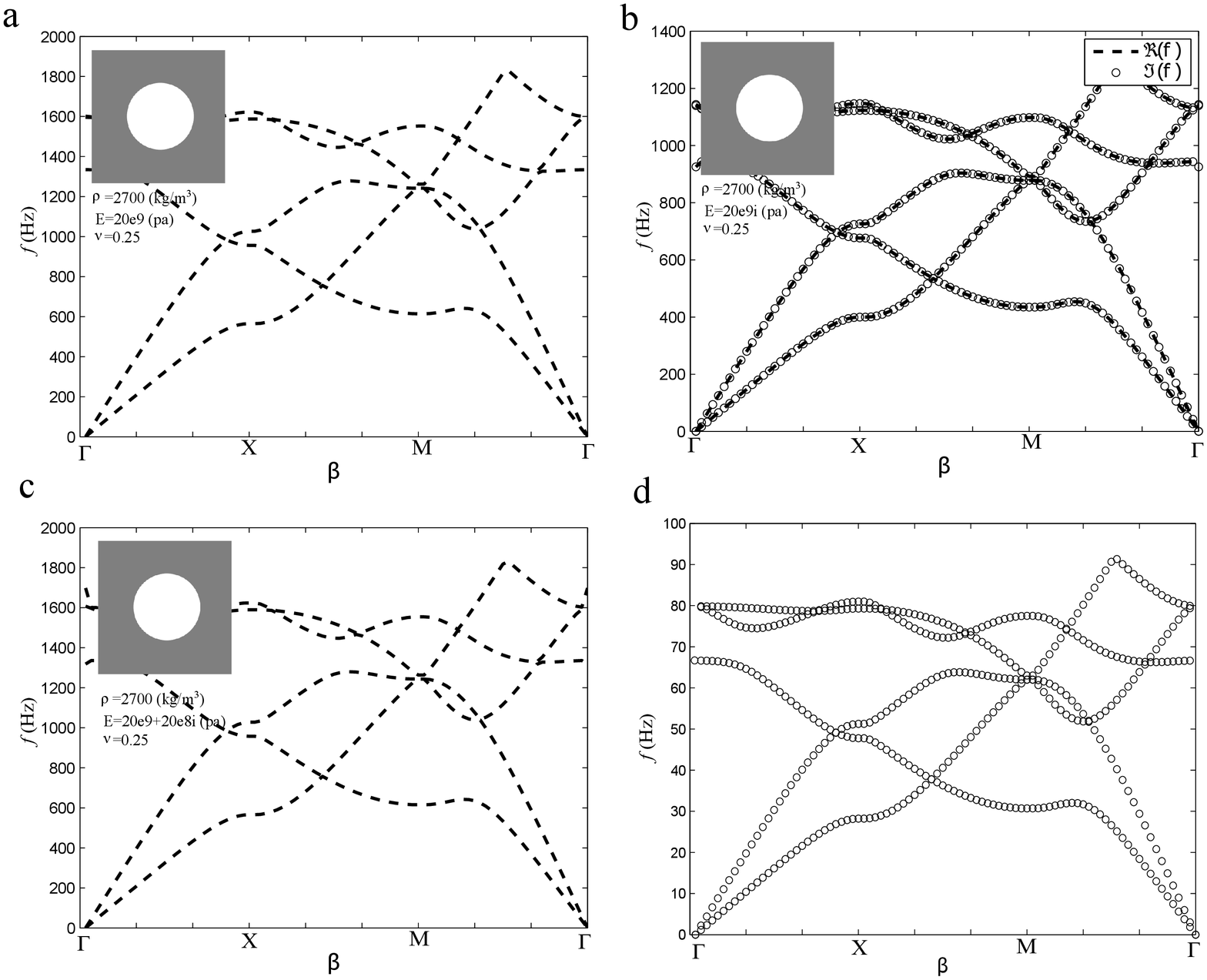}
\caption{ $\omega(\beta)$ plot for the 2D in-plane wave propagation for: (a) $\mathbf{C}=\mathbf{H}$ (b) $\mathbf{C}=\mathbf{N}$ (c) real and (d) imaginary parts of frequency  for $\mathbf{C}=\mathbf{N}+\mathbf{H}$}\label{omega-beta}
\end{figure}

The $\omega(\beta)$ solutions can be easily found by using the PWE method and the approach is standard. Consider a periodic structure with the reciprocal-lattice vectors $\mathbf{G}=(G_1,G_2,G_3)$. The material properties and field variables can be expanded using Fourier series as follows\cite{Ho1990,2013metamaterial}:
\begin{eqnarray}
\label{PWE_general1}
\alpha(\mathbf{r})=\sum_{\mathbf{G}}\alpha^{\mathbf{G}}e^{i\mathbf{G}.\mathbf{r}}
\\\label{PWE_general2} 
\mathbf{f}(\mathbf{r})=e^{-i\omega t}\sum_{\bar{\mathbf{G}}}\mathbf{f}^{\bar{\mathbf{G}}}e^{i(\bar{\mathbf{G}}+\mathbf{K}).\mathbf{r}}
\end{eqnarray}
where $\alpha$ can be any of $\{\rho, \mathbf{C},\mu, \lambda\}$, $\mathbf{f}$ can be $\{\mathbf{u},\boldsymbol{\sigma}\}$, $\mathbf{r}=(x_1,x_2,x_3)$ is the position vector, and $\mathbf{K}=\beta\mathbf{n}$ is the wave vector. Substituting the material properties and displacement field in the elastodynamics equation of motion we have:
\begin{eqnarray}
    \nabla.[\sum_{\mathbf{G}}\mathbf{C}^{\mathbf{G}}: \nabla(\sum_{\Bar{\mathbf{G}}}\mathbf{u}^{\Bar{\mathbf{G}}}e^{i(\mathbf{G}+\Bar{\mathbf{G}}+\mathbf{K})})]=
    -\omega^2\sum_{\mathbf{G}}\rho^{\mathbf{G}}\sum_{\bar{\mathbf{G}}}\mathbf{u}^{\bar{\mathbf{G}}}e^{i({\mathbf{G}+\bar{\mathbf{G}}}+\mathbf{K}).\mathbf{r}}
\end{eqnarray}
Assuming specific values of $\mathbf{K}$, the above can be solved as a generalized matrix eigenvalue problem. The calculated bandstructures for in-plane wave propagation in a square unit cell of length $1\ (m)$ with a circular hole of radius $0.25\ (m)$ is shown in Fig. (\ref{omega-beta}). In all the sub-figures, we have assumed isotopic material properties. Fig. (\ref{omega-beta}a) shows the bandstructure when $C=H$. In this case, $H$ has been defined as a self-adjoint elasticity tensor emerging from two independent material constants ($E= 20e9\ (Pa), \nu=0.25$). The frequency eigenvalues are all real, as expected from the discussion above. Fig. (\ref{omega-beta}b) shows the real and imaginary parts of the calculated frequency when $\mathbf{C}=\mathbf{N}$. In this case $\mathbf{N}$ has been defined ($E= 20e9i\ (Pa), \nu=0.25$). In this case, the real and imaginary parts of the frequencies come out to be the same, as expected by the theoretical arguments. In Figs. (\ref{omega-beta}c,d) the real and imaginary parts of the frequency are plotted when $10\%$ loss is added to the Young modulus thus giving rise to a $\mathbf{C}$ tensor which is of the form $\mathbf{H+N}$. In this case, we end up with both real and imaginary parts of the frequency eigenvalues which are not necessarily equal. 

\section{$\beta(\omega)$ solutions for a wave at a given frequency}
As a slight modification of the problem, we can seek $\beta$ solutions given frequency $\omega$ and a direction $\mathbf{n}$. The relevant eigenvalue problem can be written in several ways but we will write it in a form in which the $B$ operator, at least, has certain desirable properties. Consider the following:
\begin{eqnarray}
\displaystyle \nonumber \sigma_{ij,j}=-\omega^2\rho u_i\\
D_{ijkl}\displaystyle\sigma_{kl}=u_{i,j}
\end{eqnarray}
It will be assumed that there is minor symmetry on the compliance tensor which automatically enforces the usual requirement that $u_{i,j}=u_{j,i}$. Now considering $\mathbf{u}=\bar{\mathbf{u}}\exp(-i\beta n_ix_i)$ and $\boldsymbol{\sigma}=\bar{\mathbf{s}}\exp(-i\beta n_ix_i)$:
\begin{eqnarray}
\displaystyle \nonumber \bar{s}_{ij,j}-i\beta n_j\bar{s}_{ij}=-\omega^2\rho \bar{u}_i\\
\displaystyle D_{ijkl}\bar{s}_{kl}=\bar{u}_{i,j}-i\beta \bar{u}_{i}n_j
\end{eqnarray}
which after some rearrangement can be written as:
\begin{eqnarray}
\displaystyle \nonumber A\bar{\phi}=\lambda B\bar{\phi}
\end{eqnarray}
where $\bar{{\phi}}\equiv \{\bar{\mathbf{u}}\;\;\bar{\mathbf{s}}\}^T$, $\lambda=i\beta$ and the linear operators are given by:

\begin{eqnarray}
\nonumber \displaystyle A\bar{{\phi}}=\{\omega^2\rho\bar{u}_i+\bar{s}_{ij,j},\bar{u}_{i,j}-D_{ijkl}\bar{s}_{kl}\}\\
\displaystyle B\bar{{\phi}}=\{\bar{s}_{ij}n_j,\bar{u}_in_j\}
\end{eqnarray}

or in matrix form:

\begin{eqnarray}\label{Eq:OperA}
A=
\begin{bmatrix}
\omega^{2}\rho (\ ) & \boldsymbol{\nabla}\cdot(\ )\\
\boldsymbol{\nabla}(\ ) &-\mathbf{D}:(\ )
\end{bmatrix};\quad B=
\begin{bmatrix}
0 & (\ )\cdot \mathbf{n}\\
(\ )\otimes\mathbf{n} & 0
\end{bmatrix} 
\end{eqnarray}
In this form, $B$ is clearly self-adjoint. The adjoint operator for $A$ is:
\begin{eqnarray}
\displaystyle A^*\bar{\boldsymbol{\phi}}=\{\omega^2\rho\bar{u}_i-\bar{s}_{ij,j},-\bar{u}_{i,j}-D^*_{klij}\bar{s}_{kl}\}\\
A^*=
\begin{bmatrix}
\omega^{2}\rho (\ ) & -\boldsymbol{\nabla}\cdot(\ )\\
-\boldsymbol{\nabla}(\ ) &-\mathbf{D}:(\ )
\end{bmatrix}
\end{eqnarray}
showing that $A$ is not self-adjoint even if $D^*_{klij}=D_{ijkl}$. To check whether the operator $A$ is normal or not, we have:
\begin{eqnarray}
\nonumber
    \displaystyle  |A\bar{\phi}|^2=\langle A\bar{\phi},A\bar{\phi}\rangle=\\ 
    \displaystyle \int\left[\vert \bar{u}_{i,j}-D_{ijkl}\bar{s}_{kl} \vert^2+\vert \omega^2\rho\bar{u}_{i}+\bar{s}_{ij,j} \vert^2\right]dx
\end{eqnarray}
where $|a_i|^2=a_ia_i^*$ and $|b_{ij}|^2=b_{ij}b_{ij}^*$. Similarly we have
\begin{eqnarray}
\nonumber
    \displaystyle  |A^*\bar{\phi}|^2=\langle A^*\bar{\phi},A^*\bar{\phi}\rangle=\\ 
    \displaystyle \int\left[\vert -\bar{u}_{i,j}-D^*_{klij}\bar{s}_{kl} \vert^2+\vert \omega^2\rho\bar{u}_{i}-\bar{s}_{ij,j} \vert^2\right]dx
\end{eqnarray}
showing that in general $|A\bar{\phi}|^2\neq |A^*\bar{\phi}|^2$ and thus the operator $A$ is not normal. However, it is possible to surmise one situation in which the operator becomes normal. This would require the symmetry $D_{klij}^*=-D_{ijkl}$ and would additionally require that density be a second order tensor as well with a similar major symmetry. Specifically, $\rho_{ji}^*=-\rho_{ij}$, in addition to the aforementioned symmetry on $\mathbf{D}$ will render $A$ to be a normal operator (but not self-adjoint). In such a case we will have $A^*=-A$.

\subsection{$\beta(\omega)$ solutions using PWE}
This problem is generally a quadratic eigenvalue problem, but as shown above we can re-write it as a mixed linear eigenvalue problem. For numerical solutions we will use the following form of the eigenvalue problem:
\begin{eqnarray}
\nabla . \boldsymbol{\sigma}=\rho \ddot{\mathbf{u}}\\
\boldsymbol{\sigma}=\mathbf{C}:\nabla \mathbf{u}
\end{eqnarray}
By substituting the Fourier expansion of stress and displacement into the above equation we have:
\begin{eqnarray}
i\sum_{\hat{\mathbf{G}}}(\hat{G}_j+\beta n_j)\sigma_{ij}^{\hat{\mathbf{G}}}e^{i(\hat{\mathbf{G}}+\mathbf{K}).\mathbf{r}}=-\omega^2\sum_{{\mathbf{G}}}\rho^{\mathbf{G}}\sum_{\bar{\mathbf{G}}}u_i^{\bar{\mathbf{G}}}e^{i({\mathbf{G}+\bar{\mathbf{G}}}+\mathbf{K}).\mathbf{r}}\\
\sum_{\hat{\mathbf{G}}}\sigma_{ij}^{{\hat{\mathbf{G}}}}e^{i({\hat{\mathbf{G}}}+\mathbf{K}).\mathbf{r}}=i\sum_{\mathbf{G}}C_{ijkl}^{\mathbf{G}}\sum_{\bar{\mathbf{G}}}(\bar{G}_l+\beta n_l)u_k^{\bar{\mathbf{G}}}e^{i({\mathbf{G}+\bar{\mathbf{G}}}+\mathbf{K}).\mathbf{r}}
\end{eqnarray}
Multiplying both sides of the equations by $e^{-i({\hat{\mathbf{G}}}+\mathbf{K}).\mathbf{r}}$,  integrating over the unit cell, and separating the terms which contain $\beta$, we have the following:
\begin{eqnarray}
i\hat{G}_j\sigma_{ij}^{\hat{\mathbf{G}}}+\omega^2\sum_{{\mathbf{G}}}\rho^{\mathbf{G}}u_i^{\hat{\mathbf{G}}-\mathbf{G}}=-i\beta n_j\sigma_{ij}^{\hat{\mathbf{G}}}\\
\sigma_{ij}^{{\hat{\mathbf{G}}}}-i\sum_{\mathbf{G}}C_{ijkl}^{\mathbf{G}}(\hat{G}_l-G_l)u_k^{\hat{\mathbf{G}}-\mathbf{G}}=i\beta \sum_{\mathbf{G}}C_{ijkl}^{\mathbf{G}} n_l u_k^{\bar{\mathbf{G}}-\mathbf{G}}
\end{eqnarray}
which is the matrix eigenvalue problem whose solutions give us the wavenumber values at assumed values of frequency and wave propagation direction. 

\subsubsection{Out-of-plane waves in a 2-D phononic crystal}

\begin{figure}[htp]
\centering
\includegraphics[scale=.4]{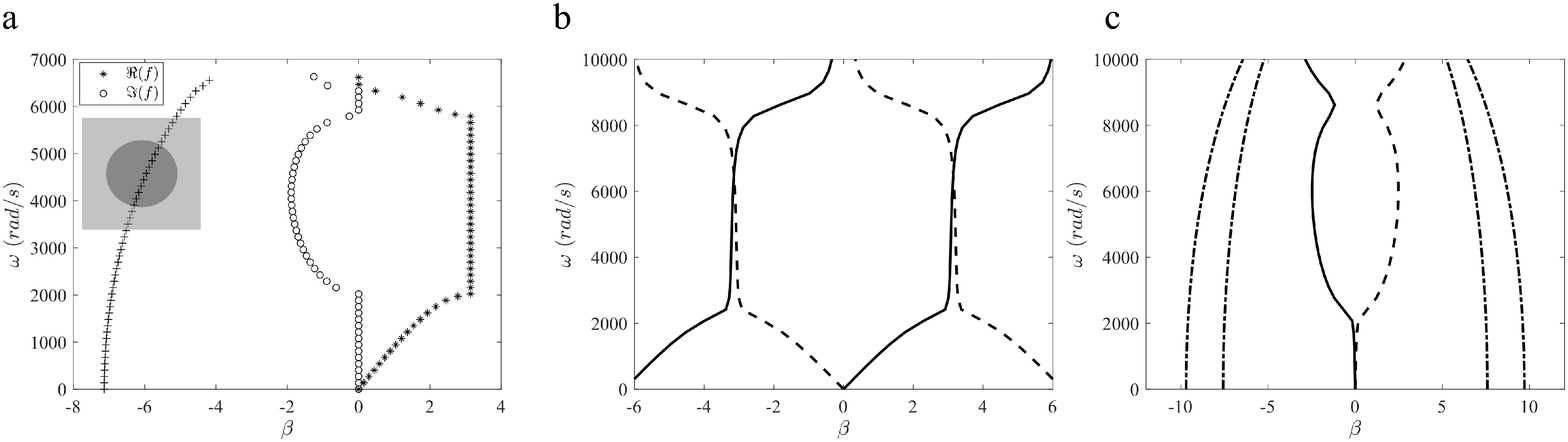}
\caption{ $\beta(\omega)$ plot for 2D out-of-plane wave propagation in the $\Gamma-X$ direction. (a) complex bandstructure for a linear elastic case. (b-c) complex bandstructure for a unit cell with a linear viscoelastic matrix and elastic  circular inclusion. In (b) the real parts of the wavevector and in (c) the imaginary parts are plotted}\label{beta-omega}
\end{figure}

To use this formulation in an example, we calculate the complex bandsructure for out-of-plane waves traveling in a square unit cell in the $\Gamma-X$ direction. The unit cell has  side of $1\ (m)$ with a central circular region of radius $0.3\ (m)$. The material properties of the matrix are $\rho_0=1100\ (kg/m^3)$, $\mu_0=1.33e9\ (Pa)$ and for the inclusion $\rho_1= 7630\ (kg/m^3)$ and $\mu_1=82e9\ (Pa)$ which are taken from Ref.\cite{Laude2009}. The complex bandstructure shown in Fig. (\ref{beta-omega}a) is for the case of linear elastic materials. In this case, only two evanescent wave mode are plotted (among infinite number of evanescent waves which exist). The first branch has zero imaginary part in the pass band and a non zero part in the band gap (where the real part is equal to $\pi$ or 0). This branch is the traditionally studied branch in the phononics literature. The second branch is purely imaginary at all frequencies and has been noted in other papers \cite{Laude2009}. It must further be noted that since this is an out-of-plane example, we do not expect complex wavenumber solutions due to the monocoupled nature of the problem and because the materials are linear elastic \cite{MEAD19751}. However, the results are still consistent with the non-normality of $A$ since the eigenvalues are $i\beta$. If this was an in-plane problem, we would expect fully complex wavenumber solutions.

The bandstructure shown in Figs. (\ref{beta-omega}b-c) are for the  unit cell of length $1\ m$ and the inclusion of radius $0.41\ m$ with the following complex shear modulus for the matrix phase (the inclusion properties are unchanged):
\begin{equation}
\mu(\omega)=\mu_0+i\eta\omega
\end{equation}
where $\eta=80\ (Pa.s)$ and $\mu_0=1.33e9\ (Pa)$. These properties are taken from Ref. \cite{Moiseyenko2011} . As the stiffness tensor is frequency dependent in viscoelastic materials, the $\omega(\beta)$ formulation cannot solve this problem directly. However, it is easy to solve this problem using the linear eigenvalue formulation of $\beta(\omega)$. In some studies\cite{Zhu2016,Zhao2009} only the real part of the shear modulus is used to compare the bandgap behavior, however in Fig. (\ref{beta-omega}-b-c) both storage and loss moduli are used. As one can see, for the viscoelastic case, pure real solutions are not predicted and the eigenvalues are complex. These results are in good agreement with Refs. \cite{Moiseyenko2011,Laude2009}.

\subsection{$\beta_3(\omega,\beta_\alpha)$ solutions}
Now we consider another kind of the phononic eigenvalue problem which one of its applications is to find the scattered wave field when a wave is incident at the interface of metamaterials \cite{srivastava2017evanescent}. In this case we are given the frequency $\omega$ and two of the three components of the wavevector $\boldsymbol{\beta}=\{\beta_1,\beta_2,\beta_3\}$. Without any loss of generality we assume that $\beta_1,\beta_2$ are known. The fields are of the form $u=\bar{u}\exp(-i\beta_ix_i)$ and $\sigma=\bar{s}\exp(-i\beta_ix_i)$. Denoting by Greek letters the indices $1,2$ and by roman the indices $1,2,3$ we can write the equation of motion as:
\begin{eqnarray}
\displaystyle \bar{s}_{ij,j}-i\beta_\alpha\bar{s}_{i\alpha}+\omega^2\rho \bar{u}_i=i\beta_3\bar{s}_{i3}
\end{eqnarray}
and the constitutive relation as:
\begin{eqnarray}
\displaystyle \nonumber -\bar{u}_{i,\alpha}+i\beta_\alpha\bar{u}_i+D_{i\alpha kl}\bar{s}_{kl}=0\\
\displaystyle -\bar{u}_{i,3}+D_{i3kl}\bar{s}_{kl}=-i\beta_3\bar{u}_i
\end{eqnarray}
These equations can be cast in the generalized eigenvalue form by identifying two new vectors. Specifically we consider $\boldsymbol{\gamma}=\{\beta_1,\beta_2,0\}$ and $\beta_3\mathbf{n}=\beta_3\{0,0,1\}$. This renders:
\begin{eqnarray}
\displaystyle \nonumber \bar{s}_{ij,j}-i\gamma_j\bar{s}_{ij}+\omega^2\rho \bar{u}_i=i\beta_3n_j\bar{s}_{ij}\\
\displaystyle \bar{u}_{i,j}-i\gamma_j\bar{u}_i-D_{ij kl}\bar{s}_{kl}=i\beta_3n_j\bar{u}_i
\end{eqnarray}
or
\begin{eqnarray}
\displaystyle \nonumber A\bar{\phi}=\lambda B\bar{\phi}
\end{eqnarray}
where $\bar{\phi}\equiv \{\bar{\mathbf{u}},\bar{\mathbf{s}}\}$, $\lambda=i\beta_3$, and $A,B$ are defined as:
\begin{eqnarray}
\nonumber \displaystyle A\bar{{\phi}}=\{\omega^2\rho\bar{u}_i-i\bar{s}_{ij}\gamma_j+\bar{s}_{ij,j},\ \bar{u}_{i,j}-i\bar{u}_i\gamma_j-D_{ijkl}\bar{s}_{kl}\}\\
\displaystyle B\bar{{\phi}}=\{\bar{s}_{ij}n_j,\bar{u}_in_j,\}
\end{eqnarray}
in matrix form:
\begin{eqnarray}\label{Eq:beta3problem}
\displaystyle A=
\begin{bmatrix}
\omega^2\rho (\ ) & \boldsymbol{\nabla}\cdot (\ )-i(\ )\cdot\boldsymbol{\gamma}\\
\boldsymbol{\nabla}(\ )-i (\ )\otimes\boldsymbol{\gamma} &-\mathbf{D}:
\end{bmatrix};\quad
\displaystyle B=
\begin{bmatrix}
0 & (\ )\cdot \mathbf{n}\\
(\ )\otimes\mathbf{n} & 0
\end{bmatrix}
\end{eqnarray}
$B$ is clearly self-adjoint. For $A$ the adjoint operator is:
\begin{eqnarray}\label{eq:Aadjoint}
\displaystyle A^*\bar{\boldsymbol{\phi}}=\{\omega^2\rho\bar{u}_i+i\bar{s}_{ij}\gamma_j-\bar{s}_{ij,j},\ -\bar{u}_{i,j}+i\bar{u}_i\gamma_j-D^*_{klij}\bar{s}_{kl}\}
\end{eqnarray}
showing that $A\neq A^*$. In general the operator is not normal either as can be seen from the following:
\begin{eqnarray}
\nonumber
    \displaystyle  |A\bar{\phi}|^2=\langle A\bar{\phi},A\bar{\phi}\rangle=\\ 
    \label{eq:Ainnerproduct}
    \displaystyle \int\left[\vert \bar{u}_{i,j}-i\bar{u}_i\gamma_j-D_{ijkl}\bar{s}_{kl} \vert^2+\vert \omega^2\rho\bar{u}_{i}-i\bar{s}_{ij}\gamma_j+\bar{s}_{ij,j} \vert^2\right]dx\\
    \nonumber
    \displaystyle  |A^*\bar{\phi}|^2=\langle A^*\bar{\phi},A^*\bar{\phi}\rangle=\\ 
    \label{eq:Aadjointinnerproduct}
    \displaystyle \int\left[\vert -\bar{u}_{i,j}+i\bar{u}_i\gamma_j-D^*_{klij}\bar{s}_{kl} \vert^2+\vert \omega^2\rho\bar{u}_i+i\bar{s}_{ij}\gamma_j-\bar{s}_{ij,j} \vert^2\right]dx
\end{eqnarray}

\subsection{PWE solution }\label{subsection:beta3PWE}
For numerical solutions we will begin with the following form of the problem:
\begin{eqnarray}
\label{interface_eq}
\displaystyle \nonumber \bar{s}_{ij,j}-i\beta_\alpha\bar{s}_{i\alpha}+\omega^2\rho \bar{u}_i=i\beta_3\bar{s}_{i3}\\
\bar{s}_{ij}-C_{ijkl}\bar{u}_{k,l}+iC_{ijk\alpha}\beta_{\alpha} \bar{u}_{k}=-iC_{ijk3}\beta_3 \bar{u}_{k}
\end{eqnarray}
By substituting Eqs. (\ref{PWE_general1},\ref{PWE_general2}) into Eq. (\ref{interface_eq}) we have the following:
\begin{eqnarray}
i\sum_{\hat{\mathbf{G}}}(\hat{G}_j+\beta_j)\sigma_{ij}^{\hat{\mathbf{G}}}e^{i(\hat{\mathbf{G}}+\mathbf{K}).\mathbf{r}}=-\omega^2\sum_{{\mathbf{G}}}\rho^{\mathbf{G}}\sum_{\bar{\mathbf{G}}}u_i^{\bar{\mathbf{G}}}e^{i({\mathbf{G}+\bar{\mathbf{G}}}+\mathbf{K}).\mathbf{r}}\\
\sum_{\hat{\mathbf{G}}}\sigma_{ij}^{{\hat{\mathbf{G}}}}e^{i({\hat{\mathbf{G}}}+\mathbf{K}).\mathbf{r}}=i\sum_{\mathbf{G}}C_{ijkl}^{\mathbf{G}}\sum_{\bar{\mathbf{G}}}(\bar{G}_l+\beta_l)u_k^{\bar{\mathbf{G}}}e^{i({\mathbf{G}+\bar{\mathbf{G}}}+\mathbf{K}).\mathbf{r}}
\end{eqnarray}
Assuming that $\beta_1$ and $\beta_2$ are given, we can re-write the above equation in the following way to form a generalized eigenvalue with $\beta_3$ as the eigenvalues and $\{\mathbf{u}, \boldsymbol{\sigma}\}$ as the eigenvectors:
\begin{eqnarray}
\label{eq:PWE_interface1}
i\sum_{\hat{\mathbf{G}}}[(\hat{G}_{\alpha}+\beta_{\alpha})\sigma_{i\alpha}^{\hat{\mathbf{G}}}+\hat{G}_3\sigma_{i3}^{\hat{\mathbf{G}}}]e^{i(\hat{\mathbf{G}}+\mathbf{K}).\mathbf{r}}+\omega^2\sum_{{\mathbf{G}}}\rho^{\mathbf{G}}\sum_{\bar{\mathbf{G}}}u_i^{\bar{\mathbf{G}}}e^{i({\mathbf{G}+\bar{\mathbf{G}}}+\mathbf{K}).\mathbf{r}}=-i\sum_{\hat{\mathbf{G}}}\beta_3\sigma_{i3}e^{i(\hat{\mathbf{G}}+\mathbf{K}).\mathbf{r}}\\
\nonumber
\label{eq:PWE_interface2}
\sum_{\hat{\mathbf{G}}}\sigma_{ij}^{{\hat{\mathbf{G}}}}e^{i({\hat{\mathbf{G}}}+\mathbf{K}).\mathbf{r}}-i\sum_{\mathbf{G}}C_{ijk\alpha}^{\mathbf{G}}\sum_{\bar{\mathbf{G}}}(\bar{G}_{\alpha}+\beta_{\alpha})u_k^{\bar{\mathbf{G}}}e^{i({\mathbf{G}+\bar{\mathbf{G}}}+\mathbf{K}).\mathbf{r}}-i\sum_{\mathbf{G}}C_{ijk3}^{\mathbf{G}}\sum_{\bar{\mathbf{G}}}\bar{G}_{3}u_k^{\bar{\mathbf{G}}}e^{i({\mathbf{G}+\bar{\mathbf{G}}}+\mathbf{K}).\mathbf{r}}=\\i\sum_{\mathbf{G}}C_{ijk3}^{\mathbf{G}}\sum_{\bar{\mathbf{G}}}\beta_{3}u_k^{\bar{\mathbf{G}}}e^{i({\mathbf{G}+\bar{\mathbf{G}}}+\mathbf{K}).\mathbf{r}}
\end{eqnarray}
multiplying both sides of the equations by $e^{-i({\hat{\mathbf{G}}}+\mathbf{K}).\mathbf{r}}$ and integrating over the unit cell, Eqs. (\ref{eq:PWE_interface1},\ref{eq:PWE_interface2}) are written as follows:
\begin{eqnarray}
i[(\hat{G}_{\alpha}+\beta_{\alpha})\sigma_{i\alpha}^{\hat{\mathbf{G}}}+\hat{G}_3\sigma_{i3}^{\hat{\mathbf{G}}}]+\omega^2\sum_{{\mathbf{G}}}\rho^{\mathbf{G}}u_i^{\hat{\mathbf{G}}-\mathbf{G}}=-i\beta_3\sigma_{i3}^{\hat{\mathbf{G}}}\\
\sigma_{ij}^{{\hat{\mathbf{G}}}}-i\sum_{\mathbf{G}}C_{ijk\alpha}^{\mathbf{G}}(\hat{G}_{\alpha}-G_{\alpha}+\beta_{\alpha})u_k^{\hat{\mathbf{G}}-\mathbf{G}}-i\sum_{\mathbf{G}}C_{ijk3}^{\mathbf{G}}(\hat{G}_{3}-G_3)u_k^{\hat{\mathbf{G}}-\mathbf{G}}=i\beta_{3}\sum_{\mathbf{G}}C_{ijk3}^{\mathbf{G}}u_k^{\hat{\mathbf{G}}-\mathbf{G}}
\end{eqnarray}
which is a generalized matrix eigenvalue problem whose solutions are $\beta_3$.

\begin{figure}[htp]
\centering
\includegraphics[scale=.7]{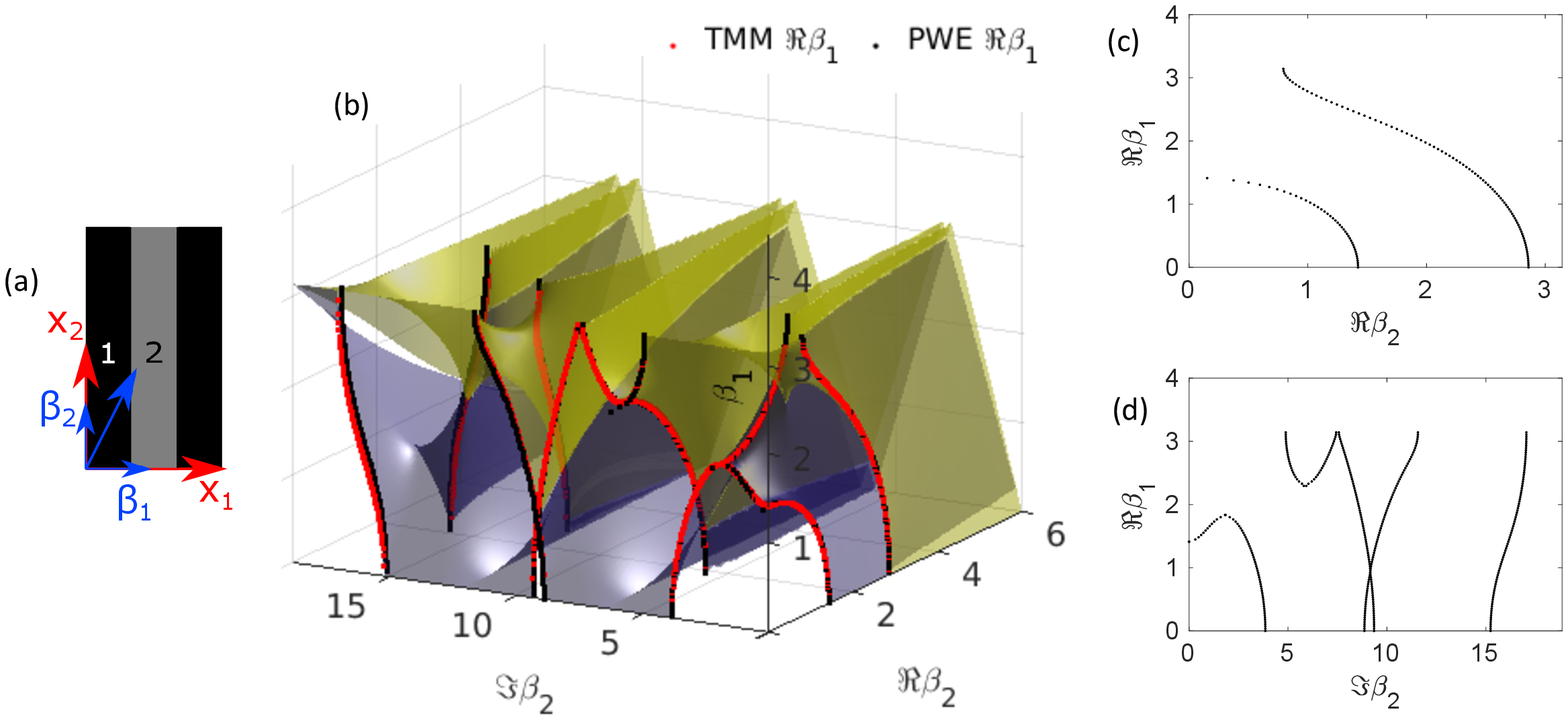}
\caption{(a) Schematic of the 1-D unit cell. (b) TMM solution for eigenvalue surface of $\beta_1$ given complex $\beta_2$ at $\omega=3000rad/s$. The two sheets of solutions are marked by yellow and blue. TMM solutions for real $\beta_1$ are marked by red dots. PWE solutions for complex $\beta_2$ given real $\beta_1$ are marked by black dots. (c) $\Re\beta_1-\Re\beta_2$ plot for the propagating waves.  (d) $\Re\beta_1-\Im\beta_2$ plot for the evanescent waves.}\label{k1-k2}.
\end{figure}

\subsubsection{In-plane waves in a 1-D phononic crystal}
To show a numerical example, we consider an in-plane wave propagation problem in periodic layered composite as an example (Fig. \ref{k1-k2}). Anti-plane shear waves in similar structures have been shown to result in exotic wave phenomenon \cite{willis2015negative,srivastava2016metamaterial} and in-plane waves have also been considered recently \cite{Shmuel2018private}. Each repeated unit cell has the following material property and layer thickness configuration:
\begin{itemize}
\item $E_1=7\ (GPa), \; \nu_1=0.25, \; \rho_1=2700\ (kg/m^3), \;  h_1=2/3\ (m),$
\item $E_2=4.34\ (GPa), \; \nu_2=0.36, \; \rho_2=1180\ (kg/m^3), \;  h_2=1/3\ (m)$
\end{itemize}

Studying some scattering problems in such a structure requires the knowledge of admissible $\beta_2$ values for given $\beta_1,\omega$ values \cite{srivastava2016metamaterial}. The numerical method of choice for solving these cases is the Transfer Matrix Method (TMM) \cite{haque2016spatial} which is an exact method. However, TMM solves for $\beta_1$ given $\beta_2,\omega$ thus requiring further numerical searches. If the eigenvalue problem is cast in the $\beta_2(\omega,\beta_1)$ form as we have done above, then it could be directly solved. Here, TMM is first employed to produce exact solutions at $\omega=3000\ (rad/s)$. At this frequency, TMM shows that there are two $\beta_1$ solutions for given $\beta_2$ values. Depending upon whether $\beta_2$ is real, imaginary, or complex, the corresponding $\beta_1$ solutions can also be complex. These two solutions are shown in Fig. (\ref{k1-k2}b) in the form of two sheets of continuous surfaces (only the real part of $\beta_1$ are plotted). Out of these multitude of solutions, there appear certain $\beta_2$ values for which $\beta_1$ is real. These real $\beta_1$ solutions are extracted from the TMM solutions using a grid search and marked by red dots in Fig. (\ref{k1-k2}b). These correspond to propagating modes in the $x_1$ direction. However, in the $x_2$ direction, these modes can be propagating (real $\beta_2$), evanescent (imaginary $\beta_2$), or non-propagating (complex $\beta_2$). This is in contrast with the anti-plane shear case \cite{srivastava2016metamaterial}, where only real or imaginary $\beta_2$ are associated with real $\beta_1$. Fig. (\ref{k1-k2}c) shows that there are two propagating modes in the $x_2$ direction for the considered range of $\beta_1$ and Fig. (\ref{k1-k2}d) shows the evanescent modes in the $x_2$ direction. In addition, there are non-propagating modes in the $x_2$ direction which are shown in Fig. (\ref{k1-k2}b) but not plotted separately. The red dots in Fig. (\ref{k1-k2}b), therefore, representing all modes which are propagating in the $x_1$ direction and can be extracted from TMM after considerable effort. These solutions are critical for solving interface problems \cite{Shmuel2018private}. Fig. (\ref{k1-k2}b) also shows the PWE solutions in black dots where the $\beta_2(\omega,\beta_1)$ problem is directly solved by varying $\beta_1$ between 0 and $\pi$. It shows that all the $\beta_2$ solutions (real, imaginary, and complex) are computed directly, as is evidenced by the comparison between the locations of the black and red dots. In the PWE solution, 73 plane waves were used in the stress and displacement expansion. 

As a further point of interest here and relating to the overall context of this work, we explicitly show the bi-orthogonality condition mentioned earlier. As shown by Eqs. (\ref{eq:Aadjoint}), (\ref{eq:Ainnerproduct}) and (\ref{eq:Aadjointinnerproduct}), the operator for the $\beta_2(\beta_1,\omega)$ problem is not self-adjoint, therefore, there is no orthogonality for right or left eigenvectors themselves. However, since there is no exceptional point in the ranges of $\beta_1,\beta_2$ considered, there is no need to create generalized eigenvectors either and the bi-orthogonality of the left and right eigenvectors should still hold. This is verified in Tab. \ref{innerproducts}. For $\beta_1=\pi/6$, $u_{(l,r)}^1$ and $u_{(l,r)}^2$ are the left and right eigenvectors, respectively, corresponding to the two $\Re\beta_2$ modes. Similarly,  $u_{(l,r)}^3$ and $u_{(l,r)}^4$ correspond to the first two $\Im\beta_2$ modes, and $u_{(l,r)}^5$ and $u_{(l,r)}^6$ correspond to the first two complex $\beta_2$ modes. The table shows that while $\Big |u_{(l)}^{i\dagger}Bu_r^j/\big|u_{(l,r)}^{i\dagger}Bu_r^i\big|\Big|=\delta_{ij}$, $\Big|u_{(r)}^{i\dagger}Bu_r^j/\big|u_{(l,r)}^{i\dagger}Bu_r^i\big|\Big|\neq\delta_{ij}$ thus showing that bi-orthogonality holds with respect to the $B$ operator.
\begin{table}[htp]
\caption{The inner products of 6 eigenmodes at $\beta_1=\pi/6$. The results below are normalized via $\Big|u_{(l,r)}^{i\dagger}Bu_r^j/\big|u_{(l,r)}^{i\dagger}Bu_r^i\big|\Big|$. Note that only the absolute values of the inner products are presented.}\label{innerproducts}
\centering
\begin{tabular}{lcccccc|cccccc}
\hline & $u_r^1$ & $u_r^2$ & $u_r^3$ & $u_r^4$ & $u_r^5$ & $u_r^6$& $u_l^1$ & $u_l^2$ & $u_l^3$ & $u_l^4$ & $u_l^5$ & $u_l^6$\\
\hline
$u_r^1$ & 1.00 & 0.77 & 0.05 & 0.01 & 0.01 & 0.01 & 1.00 & 0.00 & 0.00 & 0.00 & 0.00 & 0.00\\
$u_r^2$ & 0.19 & 1.00 & 0.17 & 0.01 & 0.05 & 0.03 & 0.00 & 1.00 & 0.00 & 0.00 & 0.00 & 0.00\\
$u_r^3$ & 0.03 & 0.41 & 1.00 & 0.15 & 0.33 & 0.05 & 0.00 & 0.00 & 1.00 & 0.00 & 0.00 & 0.00\\
$u_r^4$ & 0.12 & 0.31 & 0.13 & 1.00 & 0.20 & 0.31 & 0.00 & 0.00 & 0.00 & 1.00 & 0.00 & 0.00\\
$u_r^5$ & 0.27 & 0.66 & 0.35 & 0.29 & 1.00 & 0.04 & 0.00 & 0.00 & 0.00 & 0.00 & 1.00 & 0.00\\
$u_r^6$ & 0.05 & 0.28 & 0.19 & 0.37 & 0.05 & 1.00 & 0.00 & 0.00 & 0.00 & 0.00 & 0.00 & 1.00\\
\hline
\end{tabular}
\end{table}
\section{Conclusion}
In this paper we consider the operator properties of various phononic eigenvalue problems and aim to answer some fundamental questions about the eigenvalues and eigenvectors of phononic operators. These include questions about the potential real and complex nature of the eigenvalues, whether the eigenvectors form a complete basis, what are the right orthogonality relationships, and how to create a complete basis when none may exist at the outset. In doing so we present a unified understanding of the properties of the eigenvalues and eigenvectors which would emerge from any numerical method employed to compute such quantities. We show that the phononic problem can be cast into linear eigenvalue forms from which such quantities as frequencies ($\omega(\boldsymbol{\beta})$), wavenumbers ($\beta(\omega,\mathbf{n})$), and desired components of wavevectors ($\beta_3(\omega,\beta_\alpha)$) can be directly ascertained without resorting to searches or quadratic eigenvalue problems and that the relevant properties of such quantities can be determined apriori through the analysis of the associated operators. For the $\omega(\boldsymbol{\beta})$ problem we show that the associated operators may be normal under certain assumed symmetries of the stiffness tensor. With additional definiteness assumptions, the frequency eigenvalues can be shown to be purely real or imaginary. For the $\beta(\omega,\mathbf{n})$ and $\beta_3(\omega,\beta_\alpha)$ problems we show that the associated operators are not normal thus giving rise to complex eigenvalues in general. In each case we show how the Plane Wave Expansion (PWE) method may be extended to solve the associated eigenvalue forms and we present associated numerical examples for each eigenvalue form.

\acknowledgments     
 
A.S. acknowledges support from the NSF CAREER grant \#1554033 to the Illinois Institute of
Technology and NSF grant \#1825354 to the Illinois Institute of Technology. We would also like to thank Prof. Igor Cialenco at IIT for discussions and guidance on the spectral theorem.

%

\end{document}